\pdfoutput=1

\documentclass[11pt]{article}

\PassOptionsToPackage{table,xcdraw}{xcolor} 
\usepackage{acl}

\usepackage{times}
\usepackage{latexsym}
\usepackage{tikz}
\usepackage{relsize}
\usepackage{hyperref}
\usetikzlibrary{arrows.meta}

\usepackage[edges]{forest}
\usepackage{graphicx}
\definecolor{hidden-draw}{RGB}{0,0,0}
\usepackage{tcolorbox}
\usepackage{colortbl}
\tcbset{
  takeaway/.style={
    colback=gray!20,
    colframe=gray!70,
    boxrule=1pt,
    arc=4pt,
    left=6pt,
    right=6pt,
    top=6pt,
    title=Key Insights,
    bottom=6pt,
  }
}
\usepackage{amsfonts}       %
\usepackage{anyfontsize}
\usepackage{pgfplots}
\usepackage{multirow}       %
\usepackage{booktabs}  
\usepackage{longtable}
\usepackage{supertabular}
\usepackage{makecell}
\usepackage{pifont}
\usepackage{amsthm, amsmath}

\usepackage{todonotes}

\tcbuselibrary{skins, breakable}
\newtcolorbox{definitionbox}{
  enhanced,
  colback=gray!10,
  colframe=gray!60,
  boxrule=0.5pt,
  arc=4pt,
  left=6pt,
  right=6pt,
  top=6pt,
  bottom=6pt,
  fonttitle=\bfseries,
  title=Definition,
  breakable
}

\pgfplotsset{compat=1.15}
\usepgfplotslibrary{
  statistics,
  colorbrewer,
  groupplots,
}
\usetikzlibrary{
  patterns,
  shapes.geometric,
  decorations.text,
  matrix,
  fit,
  backgrounds,
  positioning,
}
\tikzset{
  fignode/.style={
    outer sep=0.25em,
  }
}
\tikzset{
  framedfignode/.style={
    outer sep=0.25em,
    inner sep=0.5em,
    rounded corners,
    draw,
  }
}

\usepackage[T1]{fontenc}

\usepackage[utf8]{inputenc}

\usepackage{microtype}

\usepackage{inconsolata}

\usepackage{graphicx}

\title{Test-time Corpus Feedback: From Retrieval to RAG}

\author{ Mandeep Rathee \\ L3S Research Center \\ \texttt{rathee} \\ \texttt{@l3s.de} \\
\And  Venktesh V \\ TU Delft \\ \texttt{venkyviswa12} \\\texttt{@gmail.com} \\
\And  Sean MacAvaney\\ University of Glasgow \\ \texttt{sean.macavaney} \\\texttt{@glasgow.ac.uk} \\
\And  Avishek Anand \\ TU Delft \\ \texttt{avishek.anand}\\ \texttt{@tudelft.nl}  }

\begin{document}
\maketitle

\begin{abstract}
Retrieval-Augmented Generation (RAG) has emerged as a standard framework for knowledge-intensive NLP tasks, combining large language models (LLMs) with document retrieval from external corpora. 
Despite its widespread use, most RAG pipelines continue to treat retrieval and reasoning as isolated components—retrieving documents once and then generating answers without further interaction. This static design often limits performance on complex tasks that require iterative evidence gathering or high-precision retrieval.
Recent work in both the information retrieval (IR) and NLP communities has begun to close this gap by introducing adaptive retrieval and ranking methods that incorporate feedback.
In this survey, we present a structured overview of advanced retrieval and ranking mechanisms that integrate such feedback. 
We categorize feedback signals based on their source and role in improving the query, retrieved context, or document pool. 
By consolidating these developments, we aim to bridge IR and NLP perspectives and highlight retrieval as a dynamic, learnable component of end-to-end RAG systems.

\end{abstract}

\section{Introduction}
\label{sec:intro}

Large language models (LLMs) augmented with retrieval have become a dominant paradigm for knowledge-intensive NLP tasks. In a typical \emph{retrieval-augmented generation} (RAG) setup, an LLM retrieves documents from an external corpus and conditions generation on the retrieved evidence~\cite{lewis2020rag, izacard-grave-2021-leveraging}. This setup mitigates a key weakness of LLMs—hallucination—by grounding generation in externally sourced knowledge. RAG systems now power open-domain QA~\cite{karpukhin-etal-2020-dense}, fact verification~\cite{anand2024quantemp,schlichtkrull2023averitec}, knowledge-grounded dialogue, and explanatory QA.

Despite their widespread use, many RAG systems rely on static, off-the-shelf retrieval modules—e.g., BM25~\cite{robertson1995large:bm25} or dense dual encoders~\cite{karpukhin-etal-2020-dense}—that are minimally adapted to the downstream task or domain. While re-rankers~\cite{nogueira2020document, pradeep2023rankzephyr} can improve ranking precision, the underlying retrieval often remains brittle in scenarios that demand complex reasoning: multi-hop QA, claim verification, procedural queries, or dialogue-based question answering. These tasks frequently require iterative lookups, query decomposition, or high-precision evidence—capabilities that static retrieval pipelines lack.

In contrast to the prevailing view of retrieval as a fixed first step, a growing body of work in the IR community treats retrieval as a \textit{feedback-driven, adaptive process}—where signals from the output stage is used to guide when to retrieve, how to reformulate queries, and which evidence to include.

\begin{definitionbox}
In this survey, we define \textbf{feedback} in RAGs as any signal—derived from the \textit{corpus} at different levels -- retrieval, ranking, or generation components. This feedback is used to improve the query, the context used for generation, or the set of retrieved documents.
\end{definitionbox}

We note that such feedback may be applied in one or multiple rounds and can originate from internal model signals (e.g., uncertainty or confidence), external modules (e.g., rankers or verifiers), or user behavior (e.g., clicks or clarifications).
Our notion of corpus feedback or simply feedback arises at three key stages:

\begin{enumerate}
    \item \textbf{Query-level feedback} , where the input query is rewritten, expanded, or decomposed using model introspection or relevance signals (refer Section~\ref{sec:query-understanding});
    
    \item \textbf{Retrieval-level feedback}, where rankers or corpus structure are used to revise or expand the document pool across rounds (refer Section~\ref{sec:adaptive_IR});
    \item \textbf{Generation-time feedback}, where confidence, or verifier critiques trigger new retrievals or corrections (refer Section~\ref{sec:dynamic_rag}).
\end{enumerate}

\begin{figure*}
    \centering
    \includegraphics[width=1\linewidth]{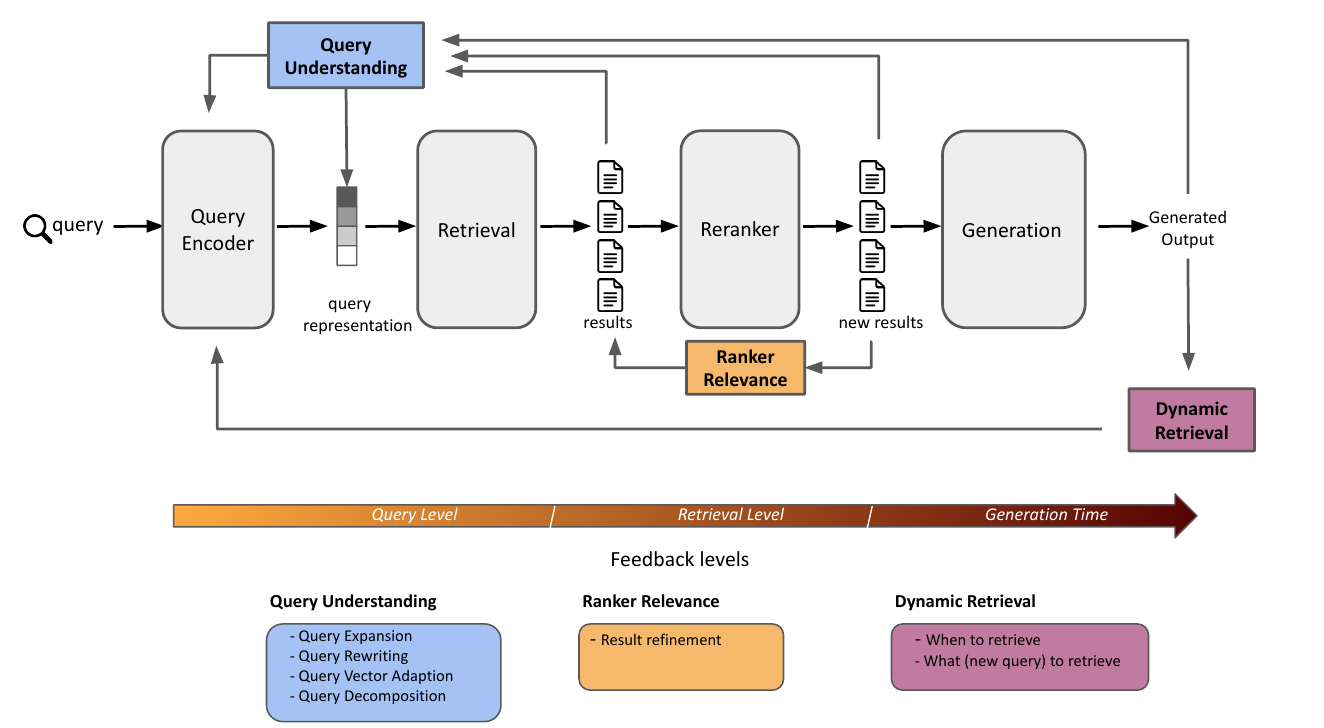}
    \caption{Illustration of feedback signals across the RAG pipeline. Feedback can modify the query (e.g., rewriting), the retrieved pool (e.g., ranker-based expansion), or the generation loop (e.g., retrieval triggers based on uncertainty).}
    \label{fig:feedback-flow}
\end{figure*}

Figure~\ref{fig:feedback-flow} shows an overview of these feedback stages. This survey synthesizes recent work that operationalizes these feedback signals across RAG pipelines. We organize methods by where and how feedback is applied -- not by architecture or dataset --emphasizing how feedback improves retrieval adaptively rather than statically.
Our scope is deliberately focused on retrieval-centric innovations in RAG. We do not cover standalone prompting or answer-generation strategies unless they directly influence the retrieval component. Our goal is to help NLP researchers treat retrieval as a dynamic, learnable component—just as vital as the generator—especially in tasks that require reasoning over incomplete, multi-part, or contextual knowledge. 
We also review the experimental landscape for retrieval-centric RAG in Section~\ref{sec:datasets}: common benchmarks, evaluation metrics, and emerging standards for assessing retriever quality in knowledge-intensive tasks.
By consolidating these developments, this survey attempts to bridge the gap between information retrieval and NLP communities, highlighting how feedback can drive the next generation of retrieval-aware, reasoning-capable RAG systems.

\section{Preliminaries}
\label{sec:preliminaries}

\subsection{Retrieval System}
\label{sec:retrieval}

The core objective of a retrieval system is to identify and rank a subset of documents $(d_1, d_2, ..., d_k)$ from a large corpus $\mathcal{C}$ based on their estimated relevance to a query $q$. Classical retrieval approaches, such as BM25~\cite{robertson1995large:bm25}, rely on exact term matching and produce sparse relevance scores. In contrast, dense retrieval methods employ neural encoders to project queries and documents into a shared embedding space, enabling semantic similarity matching~\cite{karpukhin-etal-2020-dense}.
Since first-stage retrievers often produce noisy candidates, modern pipelines incorporate a second-stage \textit{re-ranking} step using more expressive models. This includes LLM-based rankers~\cite{pradeep2023rankzephyr,rankllama2024,sun2023chatgpt} and reasoning-augmented models such as ReasonIR~\cite{shao2025reasonir}, Rank-1~\cite{weller2025rank1}, and Rank-R1~\cite{zhuang2025rank}, which refine the initial rankings by modeling deeper interactions between the query and candidate documents.

\subsection{Retrieval-Augmented Generation}
\label{sec:rag}

Retrieval-Augmented Generation (RAG)~\cite{lewis2020rag} is a hybrid paradigm that enhances the generative capabilities of large language models (LLMs) by incorporating non-parametric external knowledge during inference. This design mitigates well-documented limitations of standalone LLMs, including hallucinations, confident but incorrect outputs, and inability to reflect up-to-date or domain-specific information due to static pretraining~\cite{hurst2024gpt,grattafiori2024llama,yang2025qwen3}.

RAG introduces an explicit retrieval step: for a query $q$, a retriever selects a set of top-$k$ documents $\{d_1, \dots, d_k\}$ from an external corpus. A generator $G$ then conditions on both $q$ and the retrieved context to produce the output $a = G(q, d_1, \dots, d_k)$
where $G$ is typically an encoder-decoder or decoder-only LLM fine-tuned to integrate retrieved evidence into its generation process.

\subsection{Challenges in RAG}
\label{sec:challenges}

A central challenge in RAG is that generation quality is tightly coupled with retrieval quality. If relevant (i.e., answer-containing) documents are not retrieved, or if irrelevant context is included, the generator is prone to producing incorrect or misleading outputs~\cite{powerofnoise2024,cuconasu2025rag,liu2024lost}.

Consequently, improving the top-$k$ retrieval results is crucial. This can be viewed both as a \textit{selection} problem (how to retrieve the most relevant documents) and a \textit{filtering} problem (how to suppress distracting or noisy context). To this end, several methods have been proposed that incorporate various forms of \textit{feedback}, ranging from simple lexical overlap to more sophisticated agentic or reasoning-based signals, to guide and refine the retrieval process.

In this survey, we systematically categorize these feedback mechanisms and analyze their effectiveness across different components of the RAG pipeline. We focus on how feedback is acquired, represented, and integrated into retrieval, with the aim of providing a comprehensive taxonomy and highlighting open research challenges.

\section{Query-level feedback}
\label{sec:query-understanding}

We first focus on the first aspect which feedback in RAG systems impact -- \textit{the query}.
A fundamental factor influencing the performance of RAG systems is indeed the formulation of the input query. 
Poorly phrased, underspecified, or ambiguous queries can lead to irrelevant retrieval, ultimately degrading the quality of the generated output.

To address this, a variety of \textit{feedback-driven} query reformulation methods have been proposed. Feedback may be applied in one or multiple rounds to iteratively enhance retrieval effectiveness and overall answer quality.
In this section, we focus on the feedback improving query representations and categorize them into two broad families based on the source and representation of feedback: (i) pseudo-relevance feedback from retrieved documents, and (ii) generative relevance feedback from large language models.

\subsection{Pseudo-Relevance Feedback (PRF)}
\label{sec:lexical_prf}

Pseudo-relevance feedback (PRF) techniques modify queries based on the content of top-$k$ retrieved documents, assumed to be relevant. These methods operate either in the lexical space or in dense embedding spaces.

\paragraph{Lexical PRF.}
Classical PRF methods such as RM3~\cite{abdul2004umass}, Rocchio~\cite{rocchio1971relevance}, and KL-divergence-based models~\cite{zhai2001model} extract high-frequency terms from pseudo-relevant documents to expand the original query. These approaches rely on exact term matching and term frequency statistics. 
Enhancements like Latent Concept Expansion (LCE)~\cite{metzler2007latent} and Local Context Analysis (LCA)~\cite{xu1996query} leverage co-occurrence patterns or latent topic structures but still operate in the discrete term space. While effective for certain domains, these methods are limited by \textit{the vocabulary mismatch problem}: relevant documents may not share terms with the query, especially in low-resource or noisy scenarios.

\paragraph{Semantic PRF.}
To address lexical mismatch~\cite{EQE2016zamani,zamani2017relevance} use word embeddings to expand queries with semantically related terms. More recent techniques adopt dense retrieval settings -- ~\cite{colbert-prf2023} performs feedback-based expansion in contextualized token embedding space, while ANCE-PRF~\cite{tct_ance_prf2021} averages document embeddings to interpolate with the query vector. 
These methods enable richer semantic matching but \textit{remain sensitive to the ambiguity} or sparsity of the original query~\cite{jagerman2023query}.

\begin{tcolorbox}[takeaway]
Expands queries using information from top-ranked documents, improving recall but still limited by vocabulary mismatch, query ambiguity, and noise in initial retrievals.
\end{tcolorbox}

\subsection{Generative Relevance Feedback (GRF)}
\label{sec:grf}

Generative relevance feedback (GRF) methods employ large language models (LLMs) to generate query expansions, reformulations, or conceptually enriched representations. Unlike PRF, where feedback is extracted from retrieved documents, GRF generates feedback via prompting, generation, or learning signals.

\paragraph{LLM-only Feedback.}
Several methods prompt pre-trained LLMs to produce reformulated or expanded queries. QueryExpansion (QE)~\cite{jagerman2023query} employs different prompting styles, including Chain-of-Thought (CoT) prompting~\cite{wei2022chain}, to elicit stepwise explanations and derive new query terms. 
While these methods can function without initial retrieval, many still use retrieved documents as input, which can introduce noise. Hybrid systems such as MILL~\cite{jia2023mill}, GRF+PRF~\cite{mackie2023generativeall}, and BlendFilter~\cite{wang-etal-2024-blendfilter} combine lexical PRF and GRF by verifying consistency between generated expansions and retrieved evidence. Word-level filtering methods like Word2Passage~\cite{choi-etal-2025-word2passage} and ReAL~\cite{chen-etal-2025-terms} further refine queries using token importance estimates.

\paragraph{Feedback from Generated Answers.} Beyond generating expansions, some methods use LLM-generated answers as implicit feedback. Generation-Augmented Retrieval (GAR)~\cite{mao2020generation} generates answer-like contexts (titles, passages, summaries) using a model like BART~\cite{lewis2019bart}, which are then concatenated to the query. However, this introduces risks of hallucination and irrelevant additions. To refine this idea, RRR~\cite{arora2023gar} iteratively updates the query based on retrieval performance, using a feedback loop constrained by a document budget. LameR~\cite{shen2023large} first generates multiple answers, augments them with the query, and performs a second retrieval pass—effectively building a feedback loop from generation to retrieval. InteR~\cite{feng2024synergistic} and Iter-RetGen~\cite{shao-etal-2023-enhancing} perform tighter integration between RAG and GAR by alternating between generation and retrieval for iterative refinement. 
\paragraph{Optimization-based Feedback.}
Recent work aims to move beyond prompting heuristics by directly optimizing queries for retrieval objectives. DeepRetrieval~\cite{jiang2025deepretrieval} introduces a reinforcement learning framework where the query generation process is trained end-to-end to maximize retrieval metrics (e.g., recall, nDCG), using document-level reward signals. This eliminates reliance on manual prompting or ground truth supervision.

We refer readers to comprehensive surveys such as~\cite{song2024survey} for broader coverage of query rewriting and optimization techniques beyond the RAG context.

\begin{tcolorbox}[takeaway]
GRF methods use LLMs to generate or optimize query reformulations, offering richer semantics and adaptability, but are prone to hallucination (irrelevant but plausible-sounding terms) and require strategies to control noise.
\end{tcolorbox}

\begin{table*}
    \centering
    \resizebox{0.99\textwidth}{!}{    
    \begin{tabular}{lll}
        \toprule
        \rowcolor{gray!10}
        \textbf{Cat.1} & \textbf{Approach} & \textbf{Approach Description} \\
        \midrule
        & \textbf{Lexical Pseudo Relevance Feedback} &     \\
        \multirow{25}{*}{\rotatebox[origin=c]{90}{\textbf{Query Level}}}
        &  Lexical PRF~\citep{abdul2004umass} & Expand queries using top-k document terms  \\
        &  Rocchio~\citep{rocchio1971relevance} & Adjust vector using relevant feedback \\
        &  KL Expansion~\citep{zhai2001model} & Optimize query based on feedback documents \\
        & Adaptive Relevance Feedback~\citep{adaptive_rel_feedback2009} & Adaptive weights per query and feedback set \\
         & Relevance Modeling~\citep{metzler2005markov} & Interpolate query with new expansion terms \\
         & LCE~\citep{metzler2007latent} & Discover latent concepts for expansion   \\
         & LCA~\citep{xu1996query} & Use co-occurrence statistics for expansion  \\
        \cmidrule{2-3}
        & \textbf{Semantic Pseudo Relevance Feedback} &   \\
        & EQE~\cite{EQE2016zamani} & Words with similar embeddings are used in query expansion  \\
        & RLM/RPE~\cite{zamani2017relevance} & Train a models to output words relevance  \\
        & ANCE PRF~\citep{tct_ance_prf2021} & Expand using contrastive dense embeddings \\
        & Colbert PRF~\citep{colbert-prf2023} & Contextual embedding expansion with late interaction  \\
        \cmidrule{2-3}
        & \textbf{Generative Relevance Feedback} &    \\
        & GRF~\citep{mackie2023generativeall} & Generate contexts with LLMs for queries  \\
        & GAR~\citep{mao2020generation} & Expand using answer and passage metadata \\
        & QueryExpansion~\citep{jagerman2023query} & Prompt-based query rewriting techniques \\        
        & LameR~\citep{shen2023large} & Append generated answers to original query \\
        & InteR~\citep{feng2024synergistic} & Alternate between generation and retrieval \\
        & Iter-Retgen~\citep{shao-etal-2023-enhancing} & Interplay between GAR (or GRF) and RAG to improve answer generation  \\
        & BlendFilter~\citep{wang-etal-2024-blendfilter} & Use both LLM-generated query and original query for retrieval  \\
        & RRR~\citep{arora2023gar} & Interplay between GAR (or GRF) and RAG to improve retrieval  \\
       & MILL~\citep{jia2023mill} &Use both PRF and GRF for query expansion  \\
        & ReAL~\citep{chen-etal-2025-terms} & Learn original and expanding query terms weights   \\
        & Word2Passage~\citep{choi-etal-2025-word2passage} & Use granular word-level importance for query expansion  \\
        & DeepRetrieval~\citep{jiang2025deepretrieval} & RL training to optimize the rewritten query  \\
        \midrule
        \multirow{9}{*}{\rotatebox[origin=c]{90}{\textbf{Retrieval Level}}}
         & GraphAR~\citep{macavaney2022adaptive} & Adaptive retrieval using a corpus graph    \\
        & LADR~\citep{kulkarni2023lexically} & Use lexical results for dense retrieval   \\
        & QUAM~\citep{rathee2024quam} & Adaptive retrieval using doc-doc similarities as feedback  \\
        & LexBoost~\citep{Kulkarni2024lexboost} & Improve lexical retrieval using semantic corpus graph  \\
        & SUNAR~\citep{venktesh2025sunar} & Use answer uncertainty as feedback   \\
        & ORE~\citep{rathee:ore} & Dynamic documents selection for ranking  \\
        & SlideGAR~\citep{rathee2025guiding} & Use LLM-based listwise ranker's feedback for adaptive retrieval  \\
        & ReFIT~\citep{reddy2023refit} &Update query vector using Ranker feedback   \\
        & TOUR~\citep{sung2023optimizing} & Update query representation using ranker feedback  \\
        \midrule
        \multirow{28}{*}{\rotatebox[origin=c]{90}{\textbf{Generation-Time}}}
        & \textbf{Rule-Based Retrieval} &    \\
        & SKR~\citep{wang-etal-2023-self-knowledge} & Ask LLM if information needed    \\
        & IRCoT~\citep{trivedi-etal-2023-interleaving} & Retrieve if CoT has not provided the final answer \\
        & Adaptive RAG~\citep{jeong-etal-2024-adaptive} & Classifier's feedback for retrieval  \\
        \cmidrule{2-3}
        & \textbf{Retrieval-on-Demand via Feedback Signals} &     \\
        & FLARE~\citep{jiang-etal-2023-active} & Token probability as feedback   \\
        & DRAD~\citep{su2024mitigating} &Check hallucination in answer and trigger retrieval to mitigate   \\
        & DRAGIN~\citep{su-etal-2024-dragin} & Token probability as feedback   \\
        & Rowen~\citep{ding2024retrieve} & Answer consistency as feedback  \\
        & SeaKR~\citep{yao2024seakr} &  Internal states of the LLM as feedback  \\
        & CRAG~\cite{yan2024corrective} & Use retrieval evaluator to judge if context is relevant \\
        & CoV-RAG~\cite{he-etal-2024-retrieving} & Chain-of-Verification using a trained model  \\        
        & SIM-RAG~\cite{sim-rag2025sigir} & External critic model to judge if context is sufficient  \\
        \cmidrule{2-3}
        & \textbf{Prompt-Based Methods} &   \\
        & Self-Ask~\citep{self-ask} & Decompose the complex query into sub-queries   \\
        & DeComP~\cite{DecomP} & Decompose complex query into sub-queries   \\
        & ReAct~\cite{yao2022react} & Use each reasoning step to trigger retrieval    \\
        & Searchain~\cite{xu2024searchinthechain} & Generate chain-of-questions and trigger retrieval if needed  \\
        & MCTS-RAG~\cite{hu2025mcts} & Dynamically integrates reasoning and retrieval in MCTS   \\
        & SMR~\cite{lee2025token} & Mitigates overthinking in retrieval by guiding LLMs through discrete actions  \\
        \cmidrule{2-3}
        & \textbf{Learned or Agentic Methods} &   \\
        & Self-RAG~\cite{asai2024self} & Train LLM to predict reflection tokens that trigger retrieval and judge context  \\
      & Search-R1~\cite{jin2025search} & Train LLM to decompose query and generate tokens that trigger retrieval  \\
        & Search-O1~\cite{li2025searcho1agenticsearchenhancedlarge} & Decide autonomously when to retrieve by detecting the presence of uncertain words \\
        & R1-Searcher~\cite{song2025r1} & Reward for triggering search tokens  \\
       & ReZero~\cite{dao2025rezero} & Introduces an RL framework that rewards the act of retrying search queries \\ 
        & DeepResearcher~\cite{zheng2025deepresearcher} & Use F1 score-based reward for answer accuracy \\
        & WebThinker~\cite{li2025webthinker} & Adapt model to use commercial search engines during training  \\
       &ZeroSearch~\cite{sun2025zerosearch} & Approximate the real search engine behavior during training  \\
        \bottomrule
    \end{tabular}}
    \caption{Summary of feedback-based retrieval and RAG methods.}
    \label{tab:literature}
\end{table*}

\section{Retrieval-level feedback}
\label{sec:adaptive_IR} 

Retrieval in RAG pipelines is often bottlenecked by the bounded recall of the first-stage retriever. Once the top-$k$ documents are selected, re-ranking can improve their ordering, but cannot recover relevant documents missed in the initial retrieval. This limitation motivates \textit{adaptive retrieval} methods that incorporate feedback, often from neural rankers or structural knowledge of the corpus, to refine or expand the retrieved document set across one or more rounds. In this section, we examine two prominent classes of adaptive retrieval strategies, \textit{ neighborhood-based corpus expansion} and \textit{query vector adaptation}. 

\noindent\paragraph{\textbf{Neighborhood-based Corpus Expansion}}
relies on the \textit{clustering hypothesis} that posits that co-relevant documents tend to be similar to one another. GraphAR~\cite{macavaney2022adaptive} formalizes this intuition by constructing a corpus graph using lexical similarity between documents. After reranking an initial retrieved set, the method expands the document pool by including neighbors of top-ranked documents in the graph. Variants such as LADR~\cite{kulkarni2023lexically} and LexBoost~\cite{Kulkarni2024lexboost} improve efficiency by using dense bi-encoders and incorporating query-document and document-document edges. QUAM~\cite{rathee2024quam} generalizes these approaches by introducing query affinity modeling, taking into account the degree of similarity between neighbors and their relevance. The ORE framework~\cite{rathee:ore} further refines this strategy by prioritizing expanded documents based on their expected utility toward the ranker’s final relevance. SUNAR~\cite{venktesh2025sunar} incorporates uncertainty over multiple LLM-generated answers to adjust retrieval weights, offering a feedback loop grounded in generation uncertainty, though it may amplify hallucinated answers. SlideGAR~\cite{rathee2025guiding} uses LLM-based listwise rankers~\cite{pradeep2023rankzephyr,pradeep2023rankvicuna} to iteratively expand and refine the document pool over a document graph, closing the loop between ranking, selection, and feedback-driven retrieval.

\noindent\paragraph{\textbf{Query Vector Adaptation}} updates the query representation based on feedback from ranked documents. ReFIT~\cite{reddy2023refit} and TOUR~\cite{sung2023optimizing} both adjust the query vector in dense retrieval space using intermediate relevance scores from neural rankers. These adapted queries are used to perform second-stage retrieval, improving coverage of relevant documents.

\begin{tcolorbox}[takeaway]
Relevance feedback improves recall via efficient corpus expansion or query adaptation, but risks adding noise when similarity links or feedback are unreliable.
\end{tcolorbox}

\section{Generation-time feedback}
\label{sec:dynamic_rag}

RAG systems face two fundamental challenges: determining \textit{when to retrieve} external knowledge, since not all queries benefit from it, and \textit{how to retrieve} relevant content effectively \cite{su-etal-2024-dragin}. Classical RAG pipelines rigidly follow a fixed sequence of retrieval, optionally ranking, followed by generation, limiting their ability to adapt to the context or task.
To address these limitations, recent work has introduced \textit{ adaptive RAG}, where the retrieval strategy is dynamically adjusted according to the query, the model feedback, or the complexity of the task. We categorize this emerging line of work into three main classes.

\subsection{Rule-Based and Discriminative Approaches}
In-Context RALM (Retrieval-Augmented Language Model)~\cite{ram2023context} proposes retrieving relevant context documents during inference at fixed intervals (every $s$ tokens, known as the retrieval stride), using the last $l$ tokens of the input as the retrieval query. In a similar spirit, IRCoT (Interleaving Retrieval in a CoT)~\cite{trivedi-etal-2023-interleaving} dynamically retrieves documents if the CoT~\cite{wei2022chain} step has not provided the answer. At first, it uses the original question to retrieve the context and then uses the last generated CoT sentence as a query for subsequent retrieval. However, both of these methods retrieve the context regardless of whether the LLM needs external context or not. Hence, the unnecessary retrieval steps add additional latency cost during answer generation. Also, the noisy retrieved context can lead to a wrong answer.  
CtRLA \cite{huanshuo2025ctrla} devises a latent space probing-based approach for making decisions regarding retrieval timings for adaptive retrieval augmented generation. The authors extract latent vectors that represent abstract concepts like  \textit{honesty} and \textit{confidence} and use these dimensions to steer retrieval and control LLM behavior, leading to better performance and robust answers.
To overcome the over-retrieval limitation of rule-based dynamic RAG methods, retrieval-on-demand approaches have been proposed. These methods trigger retrieval only when the LLM needs it, based on either external feedback (Section~\ref{sec:on demand feedback sig})or the LLM’s own assessment (Section~\ref{sec:self triggered}).

\begin{tcolorbox}[takeaway]
The rule-based methods help in answer generation based on the retrieved context. However, these rules-based retrieval results in over-retrieval and add latency costs, and may provide noisy context, which can result in a wrong answer.  
\end{tcolorbox}

\subsection{Retrieval-on-Demand via Feedback Signals}
\label{sec:on demand feedback sig}
The feedback signals can come from different sources, including the answer uncertainty, the model's internal states, or context faithfulness and sufficiency. SKR~\cite{wang-etal-2023-self-knowledge} asks LLM itself if additional information is needed to answer the query. If yes, then the retrieval round is triggered; otherwise, the answer is generated from the LLM's internal knowledge. However, the judgment is solely based on LLM, and without context, they try to be overconfident~\cite{xiong2023can}. FLARE~\cite{jiang-etal-2023-active} retrieves the documents only if the token probability is below a predefined threshold and uses the last generated sentence as a query for retrieval (excluding the uncertain tokens) and generates the response until the next uncertain token or completion is done. However, these uncertain tokens are not equally important to trigger a retrieval round. Based on this, DRAD~\cite{su2024mitigating} uses an external module for hallucination detection on entities in the generated answer; if the answer contains hallucination, the retrieval is triggered. The last generated sentence (without a hallucinated entity) is used as a query for retrieval. However, the choice of the new query for retrieval relies on heuristic strategies. Since the model’s information needs may extend beyond the last sentence or CoT, it could require context from a broader span of the generation to effectively build confidence. Based on this motivation, DRAGIN~\cite{su-etal-2024-dragin}, similar to FLARE, also considers the token probabilities as a criterion of the retrieval round but does not consider the uncertain tokens as a part of the new query. Further, it also reformulates the query using the keywords based on the model's internal attention weights and reasoning. SeaKR~\cite{yao2024seakr} computes the self-aware uncertainty using internal states of the LLM. If the uncertainty is above a threshold, then a retrieval round is triggered. 

Other types of works, like Rowen~\cite{ding2024retrieve}, consider the LLM's answer consistency as feedback. Rowen considers answer consistency across languages of the same question with semantically similar variations, and the consistency over answers generated by different LLMs. If the total consistency is below a predefined threshold, then the retrieval round is triggered. However, similar to SUNAR~\cite{venktesh2025sunar}, the consistency can be toward wrong answers. 

However, these approaches consider all queries equally complex and might end up with noisy context retrieval and hence a wrong answer. Adaptive RAG~\cite{jeong-etal-2024-adaptive} uses a query routing mechanism that predicts whether the query needs retrieval or not. Further, it also decides on the number of retrieval rounds based on query complexity. However, it assumes that the retrieved context is relevant to the query without assessing its relevancy or sufficiency. Towards the relevancy, CRAG (Corrective RAG)~\cite{yan2024corrective} evaluates the relevance scores using a fine-tuned model, and classifies the retrieved document into correct, incorrect, and ambiguous. If the context is not correct, then a rewritten query is issued to the web search engine. Similar fashion, SIM-RAG~\cite{sim-rag2025sigir} focuses on the context sufficiency angle, and trains a lightweight critic model that provides feedback if the retrieved context is sufficient to generate the answer. If the information is not sufficient, then a new query is formulated using the original query and the already retrieved context, and a retrieval round is triggered. Further CoV-RAG~\cite{he-etal-2024-retrieving} identifies errors, including reference and answer correctness, and truthfulness, and then scores them using a trained verifier. Based on the scores, either provide a final or rewrite the query and do a further retrieval round. 

\begin{tcolorbox}[takeaway]
The external feedback signals help in reducing retrieval rounds. These signals can come from different sources, at the LLM level (e.g., token generation confidence), at the answer level ( e.g., uncertainty or hallucination), and at the context level (e.g., relevancy or sufficiency). 
However, these methods may still retrieve noisy or irrelevant context, and complexity assessment remains a challenge.
\end{tcolorbox}

\subsection{Self-Triggered Retrieval via Reasoning}
\label{sec:self triggered}
In this section, we discuss works where LLM autonomously makes the decision on when to retrieve and how to retrieve through query decomposition or planning-based approaches without external triggers. These approaches are also termed \textit{Reasoning RAG} or \textit{Agentic RAG}. These approaches can be divided into mainly two categories: first, where the instructions for query decomposition, when to retrieve, and what to retrieve are provided in the prompt along with few-shot examples; second, where the language models are trained to decide by themselves whether to decompose the query, when to retrieve, and what to retrieve. 

\paragraph{Prompt-Based Methods.}
DeComP \cite{DecomP} divides a task into granular sub-tasks and delegates them to different components through actions. However, DeComP only acts as a trigger for when to retrieve and employs a BM25  retriever for getting relevant documents in a single shot. It does not subsequently generate reasoning steps to improve retrieval, thus not providing much indication as to how to retrieve. ReAct \cite{yao2022react} interleaves the generation of verbal reasoning traces with actions that interact with the external environment. The verbal reasoning traces act as indicators of how to retrieve, and the actions themselves serve as triggers (when to retrieve). Similarly, Self-Ask \cite{self-ask} proposes to decompose the original complex query into simpler sub-questions iteratively interleaved by a retrieval step and intermediate answer generation. At each step, the LLM makes a decision to generate a follow-up question if more information is needed, or it may generate the final answer. Authors observed that this approach helped cover diverse aspects of complex queries and improved search and downstream answering performance.

However, these approaches do not have provision for correction of the entire reasoning trajectory, and an intermediate error may cause cascading failures. Searchain \cite{xu2024searchinthechain} proposes to mitigate this by constructing a global reasoning chain first, where each node comprises a retrieval-oriented query, an answer from LLM to the query, and a flag indicating if additional knowledge is needed to arrive at a better answer. SMR (State Machine Reasoning)~\cite{lee2025token} identifies the issues of the CoT-based query decomposition and retrieval methods like ReAct~\cite{yao2022react}, where the CoT might result in redundant reasoning (new queries that result in the retrieval of the same documents) and misguided reasoning (new query diverges from the user's intent). To address these limitations, SMR proposes three actions: Refine, Rerank, and Stop. Action Refine updates the query using the feedback from the already retrieved documents, and a retrieval round is triggered. Then the retrieved documents are ranked according to the old query to make sure only the relevant information is used to answer. Finally, the Stop action is called to stop the reasoning if a sufficient retrieval quality is achieved, which helps in token efficiency and prevents overthinking. 

MCTS-RAG~\cite{hu2025mcts} combines Monte Carlo Tree Search (MCTS) with Retrieval-Augmented Generation (RAG) to improve reasoning and retrieval in language models. It guides the search for relevant information using MCTS to explore promising retrieval paths, enhancing answer accuracy. However, it is computationally expensive due to the iterative tree search process and may struggle with highly noisy or irrelevant documents. Search-O1~\cite{li2025searcho1agenticsearchenhancedlarge} proposes an agentic search workflow for reasoning augmented retrieval by letting the Large Reasoning Models (LRMs) like O1 decide autonomously when to retrieve by detecting the presence of salient uncertain words in their output. Additionally, they augment the workflow with a reason-in-documents step, where LRMs analyze the documents in depth to remove noise and reduce redundancy before employing them to generate the final answer. 

\begin{tcolorbox}[takeaway]
Query decomposition and interleaving reasoning with retrieval improve coverage for complex questions by deciding when and what to retrieve, but are prone to cascading errors and redundant steps.
\end{tcolorbox}

\paragraph{Learned or Agentic Methods.}

The Agentic models go beyond prompt instructions and use search/retrieval as a tool. These models are trained to trigger this tool during answer generation. The training process mainly focuses on giving rewards for correct tool calls and context usage. In addition, similar to RAG methods, the retrieved documents are used as context to generate intermediate answers or the final answer. The search tool might have access to a local database or a web search engine to retrieve up-to-date knowledge. 

Self-RAG~\cite{asai2024self} trains to predict reflection tokens for deciding when to retrieve and for estimating the relevance of retrieved documents. In addition, it judges the retrieved documents based on the generated answers and their factuality. However, it can fail when its self-reflection misjudges retrieval needs or relevance, leading to missed information or reliance on irrelevant context. 

Search-R1~\cite{jin2025search} is an extension of the DeepSeek-R1~\cite{guo2025deepseek} model, where the retrieval is a component training process. It autonomously generates search queries and performs real-time retrieval during step-by-step reasoning processes through reinforcement learning, including GRPO and PPO. The retrieval is triggered by \texttt{<search>} and \texttt{</search>} tokens, and the retrieved context is enclosed in \texttt{<information>} and \texttt{</information>} tokens. Similarly, R1-Searcher~\cite{song2025r1} also uses an RL framework and uses two-stage rewards. The first stage has a retrieval reward that helps the model to use the correct format to trigger the retrieval, and the second stage has an answer reward that encourages the model to learn to utilize external retrieval effectively. While both these methods encourage better integration of external knowledge, they still inherit the limitations of retrieval latency and potential noise from the search source.

ReZero (Retry-Zero)~\cite{dao2025rezero} introduces an RL framework that rewards the act of retrying search queries following an unsuccessful initial attempt, and it encourages LLM to explore alternative queries rather than prematurely stopping. The training process provides positive signals/rewards (feedback) if the model executes a retry action after failed searches, teaching the philosophy of "try one more time". However, these local database-based searches might miss the up-to-date knowledge and could generate answers for queries that require such knowledge. 

DeepResearcher~\cite{zheng2025deepresearcher} and WebThinker~\cite{li2025webthinker} interact in real time with commercial search engines during training, which leads to noisy context from the web (since the quality of these documents is unpredictable) and a high number of API calls. To address these limitations, ZeroSearch~\cite{sun2025zerosearch} argues that since LLM has acquired enough world knowledge during heavy pre-training, it does not need to use a search engine during training. Since the LLM itself can generate a good-quality document from its parametric memory that answers the query, as well as noisy documents. Hence, it can approximate the real search engine behavior during training and reduce the training costs and noise, but its effectiveness depends on the LLM's pre-trained knowledge. 

\paragraph{Verifier-Based Feedback.} Re$^2$Search++~\cite{raggym} proposes a fine-tuned critic model that verifies intermediate answers and provides feedback based on its correctness to improve the quality of intermediate queries and retrieval.

\begin{tcolorbox}[takeaway]
Learned or Agentic methods train models to decide when and how to retrieve, boosting autonomy and integration, but introduce retrieval latency, noise, and dependence on external web search. In addition, the use of an external web search engine makes it difficult to evaluate the retrieval performance. 
\end{tcolorbox}

\section{Datasets and Evaluation Benchmarks}
\label{sec:datasets}

The evaluation of IR and RAG systems relies on diverse datasets that test different aspects of retrieval and answer generation capabilities, including the retrieval, ranking, and answer quality. 

\subsection{IR Specific}
Information retrieval has a long history of evaluation campaigns, including those from TREC, CLEF, NTCIR, and FIRE. The queries used in these collections are often developed to strike a balance between having too many relevant documents in the target collection (which can be too easy to retrieve and too difficult to properly annotate~\cite{DBLP:conf/sigir/VoorheesCL22}) and too few relevant documents. Sometimes challenging topics are also developed deliberately~\cite{DBLP:journals/sigir/Voorhees05}. The topic development process often involves manual reformulation of queries to ensure good coverage of relevance assessments. Mirroring the annotation process itself, it can be beneficial for automated retrieval systems to also perform various forms of query understanding (expansion or rewriting) to help ensure high recall. Hence, the approaches described in Section~\ref{sec:query-understanding} show performance gains. 

Comprehensive benchmarks like BEIR~\cite{thakur2021beir} offer heterogeneous evaluation across 17 diverse datasets spanning multiple domains and retrieval tasks, enabling zero-shot generalization assessment. In addition, evaluation sets such as TREC DL (Deep Learning) and its variants~\cite{craswell2020overview,craswell2021overview,craswell2022overview,craswell2023overview} are used to evaluate IR systems.  

\subsection{Question Answering}
Question Answering datasets for retrieval evaluation fall into two primary categories based on complexity. Single-hop QA benchmarks include Natural Questions (NQ)~\cite{kwiatkowski-etal-2019-natural}, TriviaQA~\cite{joshi-etal-2017-triviaqa}, SQuAD~\cite{rajpurkar2016squad}, and PopQA~\cite{mallen2022not} test the RAG system's ability to retrieve and utilize information to generate the final answer. Multi-hop QA benchmarks comprise 2WikiMultiHopQA~\cite{ho-etal-2020-constructing} (questions requiring evidence from exactly two Wikipedia articles via bridging/comparison), HotpotQA~\cite{yang-etal-2018-hotpotqa}, and MuSiQue~\cite{trivedi2022musique}, which features compositionally complex questions constructed from interconnected sub-questions. Due to the dependency on intermediate answers, such questions cannot be answered through isolated single-step retrieval. Therefore, this design naturally motivates the RAG systems to decompose the original query and new queries based on the intermediate answers help in the retrieval of better context.

\subsection{Fact Checking/Verification} These datasets assess models' ability to verify claims against retrieved evidence. FEVER (Fact Extraction and VERification)~\cite{thorne2018fever}  is an open-domain fact-checking benchmark that requires retrieval over a large collection of Wikipedia articles and training NLI models to classify claims as supported, refuted, or not having enough information based on Wikipedia evidence. While FEVER deals with simple claims, HoVeR \cite{jiang-etal-2020-hover} proposes a fact-checking benchmark with claims that require multi-hop reasoning and multi-turn retrieval. QuanTemp~\cite{anand2024quantemp} is the first to propose a large open-domain benchmark for fact-checking numerical claims. It comprises claims that require interleaving claim decomposition and retrieval based on the verification output of sub-claims. This relates to query-level feedback and generation-time feedback. AveriTeC~\cite{schlichtkrull2023averitec} is a real-world claim verification dataset that includes diverse claim types with supporting or refuting evidence gathered through web search, making it particularly valuable for evaluating RAG systems on authentic fact-checking scenarios that require verification against potentially noisy or contradictory web sources.

\subsection{Complex Reasoning}
Emerging benchmarks introduce novel relevance criteria requiring awareness of reasoning structure. BRIGHT~\cite{su2024bright} defines document relevance not by topical alignment but by whether passages contain \textit{logical constructs} (deductive steps, analogies, constraints) necessary to derive answers. This challenges lexical retrievers that lack inference-awareness. 
However, reasoning-augmented retrieval methods, such as CoT (Chain-of-Thought), have shown performance gains. Further, there exist reasoning/agentic tasks like Deep Research~\cite{wu2025agentic}, GPQA~\cite{gpqa}, MATH500~\cite{cobbe2021trainingverifierssolvemath}, which involve access to the web search engine, have been used in the RAG setting~\cite{li2025searcho1agenticsearchenhancedlarge,browsecomp2025wei}. However, due to the absence of the corpus, it is hard to evaluate the retrieval performance. Recent benchmarks like BrowseComp-Plus~\cite{chen2025BrowseCompPlus} provide a curated corpus for Deep Research tasks and enable the evaluation of retrieval performance.

\subsection{Evaluation Metrics}
Evaluating retrieval systems and Retrieval-Augmented Generation (RAG) pipelines is critical for ensuring the accuracy, relevance, and reliability of generated responses. Retrieval evaluation typically focuses on metrics such as recall@k, precision, mean reciprocal rank (MRR), and hit rate, which assess how effectively the system retrieves pertinent documents or passages given a query. In contrast, RAG evaluation is more holistic, combining retrieval quality with generation fidelity and coherence. Common approaches include measuring answer correctness using exact match (EM) or F1 score, assessing faithfulness to retrieved evidence to detect hallucinations, and evaluating relevance and fluency through human or automated scoring (e.g., BLEU, ROUGE, or BERTScore).
Recent frameworks like RAGAS~\cite{ragas2024}, ARES~\cite{saad2023ares}, and CRUX~\cite{ju2025controlled} also emphasize end-to-end evaluation, where the interplay between retrieval accuracy and generation quality is analyzed to identify bottlenecks—such as irrelevant documents leading to incorrect answers—making comprehensive evaluation essential for diagnosing and improving RAG system performance.

\noindent\textbf{Challenges}. The current RAG evaluation methods mainly focus on the retrieval and final answer performance. However, Reasoning RAG systems are highly dependent on intermediate reasoning steps and retrieval rounds. Therefore, it is also important to consider additional evaluation dimensions such as computational cost, efficiency, or number of retrieval rounds.

\section{Challenges and Future Directions}
\label{sec:challenges}

Despite recent advances, test-time corpus-level feedback in RAG systems faces several key limitations related to computational cost, feedback quality, decision-making, and evaluation.

\paragraph{Computational Cost of Adaptive Retrieval.}
Many feedback-driven approaches involve costly operations such as multiple retrieval rounds, re-ranking with large models, or traversing corpus graphs~\cite{rathee2024quam, kulkarni2023lexically, hu2025mcts}. These methods often apply uniformly across queries, regardless of complexity. Efficient strategies—e.g., lightweight rankers, selective triggering, or confidence-aware stopping~\cite{jiang-etal-2023-active, yao2024seakr}—are crucial to make such systems viable at scale. Notably, recent work shows that smaller models can achieve competitive performance if given high-quality context~\cite{venktesh2025sunar}, highlighting the importance of retrieval efficiency.

\paragraph{Noisy and Unstructured Corpus Feedback.}
Retrieved documents often contain redundant or irrelevant content, and most systems lack mechanisms to assess document utility beyond relevance ranking. Few methods exploit inter-document structure such as semantic similarity or topical diversity~\cite{macavaney2022adaptive, rathee2024quam}. Structured representations (e.g., retrieval graphs, clusters) could improve feedback signals by enabling more targeted document selection and filtering.

\paragraph{Lack of Feedback-Aware Decision Policies.}
Many RAG systems perform fixed sequences of retrieval and reformulation without explicit decision criteria for when feedback is sufficient or which action (re-ranking, rewriting, reretrieval) to take~\cite{su-etal-2024-dragin, li2025searcho1agenticsearchenhancedlarge}. Learning retrieval control policies based on document-level or generation-time signals is a promising but underexplored direction.

\paragraph{Inadequate Evaluation of Feedback Behavior.}
Existing benchmarks emphasize answer correctness or static retrieval recall, but rarely measure feedback effectiveness across rounds. Datasets often lack annotations for document utility, retrieval iteration, or evidence sufficiency. Metrics that credit systems for improving retrieval through feedback—e.g., via answer change, document set refinement, or reduced over-retrieval—are needed to advance the field~\cite{zheng2025deepresearcher}.

\medskip
\noindent 

We believe that tackling these challenges is essential to make corpus-level feedback a robust and efficient component of real-world RAG pipelines, closing the loop between retrieval and reasoning in complex language tasks.

\section*{Limitations}
\label{sec:limitations}

This survey focuses exclusively on \textbf{test-time feedback mechanisms that involve interaction with the corpus} in Retrieval-Augmented Generation (RAG) systems. We refer to this as \textit{corpus-level feedback}—signals derived from retrieved documents, re-rankers, document-document relationships, or other corpus-grounded structures. Several related forms of feedback fall outside our scope.
First, we do not cover feedback mechanisms that operate independently of the corpus—such as LLM self-refinement, planning, or reasoning without retrieval. For example, techniques that rewrite queries based solely on model introspection (e.g.,  self-refine~\cite{madaan2023selfrefineiterativerefinementselffeedback}) without consulting retrieved content are not considered corpus feedback and are excluded.

Second, our focus is restricted to retrieval-centric adaptation. 
We do not survey approaches that modify the generation module unless they directly inform or adapt retrieval via corpus-level signals.
Third, we do not cover training-time feedback or methods that rely on offline supervised signals to fine-tune retrievers. Our interest is in test-time feedback mechanisms that dynamically update the query, document pool, or ranking without modifying model parameters.

\bibliography{reference}
\clearpage
\appendix

\section{Literature Compilation}

\subsection{Search Strategy}
We conducted a comprehensive search on Google Scholar. We first focused on highly relevant Natural language Processing (NLP) venues such as ACL, EMNLP, NAACL, COLM and journals like TACL to collect RAG related literature. We also extensively curated papers from IR venues like SIGIR, ECIR, CIKM, WSDM to cover information retrieval literature and recent advancements in RAG systems. 

\subsection{Compilation Strategy}

After careful review of Abstract, Introduction, Conclusion and Limitations we only retained papers that employ feedback mechanisms for improving retrieval and other components of RAG system which also helped synthesize our definition of feedback described in Section \ref{sec:query-understanding}

\end{document}